\documentclass[letterpaper, 10 pt, conference]{ieeeconf}  % Comment this line out if you need a4paper

\usepackage[T1]{fontenc}
\usepackage[utf8]{inputenc}
\IEEEoverridecommandlockouts                              % This command is only needed if 
\usepackage{graphicx}      % include this line if your document contains figures
\usepackage{hyperref}
\usepackage{tabularx}
\usepackage{caption}
\usepackage{amsmath}
\usepackage{mathtools}
\usepackage{siunitx}
\usepackage{subfigure}
\usepackage{algorithm}
\usepackage{booktabs}
\usepackage{url}
\usepackage{cite}
\usepackage{amsmath,amssymb,amsfonts}
\usepackage{algorithmic}
\usepackage{graphicx}
\usepackage{textcomp}
\usepackage{xcolor}
\usepackage{float}
\def\BibTeX{{\rm B\kern-.05em{\sc i\kern-.025em b}\kern-.08em
    T\kern-.1667em\lower.7ex\hbox{E}\kern-.125emX}}

\newcommand{\ui}[2]{#1_{\text{#2}}}

\newcommand{\Ts}{\ui{T}{s}}

\definecolor{change}{RGB}{255,0,0}      % red for "changes marked"

\definecolor{koopmanpurple}{HTML}{7030A0}  % Dark purple from plots
\definecolor{n4sidblue}{HTML}{4A90E2}      % Steel blue from plots
\definecolor{koopmanred}{HTML}{8B0000}     % Dark red (nonlinear Koopman)
\definecolor{constraintred}{RGB}{255,0,0}   % Red zone
\definecolor{softorange}{RGB}{255,165,0}    % Orange zone
\definecolor{desiredgreen}{RGB}{0,128,0}    % Green zone

\title{\LARGE \bf
Deep Koopman Economic Model Predictive Control \\of a Pasteurisation Unit
}

\author{Patrik Val\'abek$^{1}$, Michaela Horv\'athov\'a$^{1}$, and Martin Klau\v{c}o$^{1,2}$% <-this % stops a space
\thanks{The authors M. Klaučo and P. Valábek gratefully acknowledge the contribution of the Scientific Grant Agency of the Slovak Republic under the grants VEGA 1/0239/24, the Slovak Research and Development Agency under the project APVV-20-0261. P. Valábek is supported by an internal STU grants for young researchers and acknowledges the contribution of the EIT Manufacturing – Slovakia – X Fund. M. Klaučo is also supported by the European Union project ROBOPROX (Reg. No. CZ.02.01.01/00/22\_008/0004590). M. Horváthová acknowledges the contribution of the EU NextGenerationEU through the Recovery and Resilience Plan for Slovakia under the project No. 09I03-03-V04-00636.}% <-this % stops a space
\thanks{$^{1}$Slovak University of Technology in Bratislava, Institute of Information Engineering, Automation, and Mathematics, Bratislava, Slovakia}%
\thanks{$^{2}$Czech Technical University, Department of Control Engineering, Prague, Czechia}%
}

\begin{document}

\maketitle
\thispagestyle{empty}
\pagestyle{empty}

%%%%%%%%%%%%%%%%%%%%%%%%%%%%%%%%%%%%%%%%%%%%%%%%%%%%%%%%%%%%%%%%%%%%%%%%%%%%%%%%
\begin{abstract}
    This paper presents a deep Koopman-based Economic Model Predictive Control (EMPC)  for efficient operation of a laboratory-scale pasteurization unit (PU). The method uses Koopman operator theory to transform the complex, nonlinear system dynamics into a linear representation, enabling the application of convex optimization while representing the complex PU accurately. The deep Koopman model utilizes neural networks to learn the linear dynamics from experimental data, achieving a 45\% improvement in open-loop prediction accuracy over conventional N4SID subspace identification. Both analyzed models were employed in the EMPC formulation that includes interpretable economic costs, such as energy consumption, material losses due to inadequate pasteurization, and actuator wear. The feasibility of EMPC is ensured using slack variables. The deep Koopman EMPC and N4SID EMPC are numerically validated on a nonlinear model of multivariable PU under external disturbance. The disturbances include feed pump fail-to-close scenario and the introduction of a cold batch to be pastuerized. These results demonstrate that the deep Koopmand EMPC achieves a 32\% reduction in total economic cost compared to the N4SID baseline. This improvement is mainly due to the reductions in material losses and energy consumption. Furthermore, the steady-state operation via Koopman-based EMPC requires 10.2\% less electrical energy. The results highlight the practical advantages of integrating deep Koopman representations with economic optimization to achieve resource-efficient control of thermal-intensive plants.
\end{abstract}

\begin{keywords}
Koopman operator theory, economic model predictive control, data-driven control, system identification, neural networks, pasteurization, thermal process control
\end{keywords}

%%%%%%%%%%%%%%%%%%%%%%%%%%%%%%%%%%%%%%%%%%%%%%%%%%%%%%%%%%%%%%%%%%%%%%%%%%%%%%%%
\section{INTRODUCTION}
Efficient operation of thermal-intensive plants, such as pasteurization units (PU), is a challenge, as safety, product quality, material waste, and energy consumption must be balanced simultaneously. Conventional control approaches rely on first-principles models, which are often too simplistic due to strong nonlinearities in heat transfer and fluid dynamics~\cite{MARTIN2018}. Model Predictive Control (MPC) is a suitable strategy for such multivariable systems due to its explicit constraint handling and optimal performance. However, classical MPC requires a computationally tractable prediction model, typically linear, which degrades control performance in highly nonlinear plants~\cite{MAYNE2000789}.

Koopman operator theory~\cite{koopman1931hamiltonian,mezic2005spectral} provides a powerful mathematical framework to represent nonlinear dynamics as linear evolution in a lifted space of observables. Its data-driven approximations, such as Dynamic Mode Decomposition (DMD)~\cite{schmid2010dynamic} and Extended DMD~\cite{williams2015data}, have attracted attention for enabling system identification without explicit physical modelling. Recent works have extended these methods using neural networks, allowing the lifting and projection mappings to be learned directly from data~\cite{korda2018linear,yeung2019learning}. These deep Koopman formulations retain linear dynamics in the lifted space, making them particularly suitable for integration with MPC frameworks~\cite{valabek2025deep,korda2020optimal}.

At the same time, Economic Model Predictive Control (EMPC)~\cite{ellis2017economic,rawlings-empc} generalizes the conventional MPC formulation by directly optimizing an economic objective rather than a tracking error, enabling operation at economically optimal steady-states. EMPC has proven valuable in energy-intensive applications where efficiency and resource utilization are more important than strict reference tracking.

Our paper combines these two frameworks by developing a deep Koopman-based EMPC for an experimental pasteurization unit (PU). The idea of integrating Koopman representations with EMPC objectives has already been explored in several recent studies. Some studies have proposed reinforcement learning of Koopman models optimized directly for control performance within economic nonlinear MPC formulations~\cite{empc-cstr}. This approach learns the Koopman representation and the control objective at the same time, improving closed-loop performance on the benchmark of nonlinear reactors. Work by~\cite{empc-wastewater} proposed a deep-learning-based Koopman model for wastewater treatment and employed a convex EMPC formulation to improve the economic performance of the plant. In ~\cite{empc-ded}, the linear Koopman model replaced the nonlinear generator dynamics. Enabling a convex optimal control problem and using a neural policy to be learned offline. This removed the need for online nonlinear optimization, achieving considerable computational savings. Although this formulation optimizes an economic objective, it does not employ EMPC in the receding-horizon sense, it replaces online optimization with an offline-trained Koopman-based controller. Extended DMD was considered to obtain a Koopman-based linear model of a nonlinear wave-energy converter suitable for EMPC design in~\cite{empc-wave}.
The Koopman-based EMPC maximized power extraction and ensured asymptotic stability of the closed-loop system, demonstrating the effectiveness of Koopman representations for economic optimization of energy-intensive systems.

The linear-projection deep Koopman model in this paper is based on experimental data from a laboratory-scale PU. The model is compared to a conventional subspace identification (N4SID) model under identical disturbance scenarios. The obtained linear models are compared within a tailored EMPC formulation.
The EMPC explicitly incorporates an economic stage cost reflecting actual energy consumption, expected material loss, and expected actuator wear. The optimal control problem includes soft input and output constraints implemented through physically interpretable slack variables to achieve feasibility under disturbances. Numerical results on a nonlinear-projection Deep Koopman model demonstrate that the Koopman-based EMPC achieves up to a 32\% reduction in the EMPC cost function compared to conventional N4SID identification. The reduction is achieved by lower material losses and lower energy consumption, showing the potential of Koopman representations for practical, resource efficient process control.

\section{Preliminaries}
The following section recalls the Koopman operator theory and data-driven modelling approach considered in this paper. The formulation of a general Economic Model Predictive Control with slack variables is also introduced.
\subsection{Deep Koopman-Based System Identification}

\begin{figure}[h]
\centering
\includegraphics[width=0.8\columnwidth]{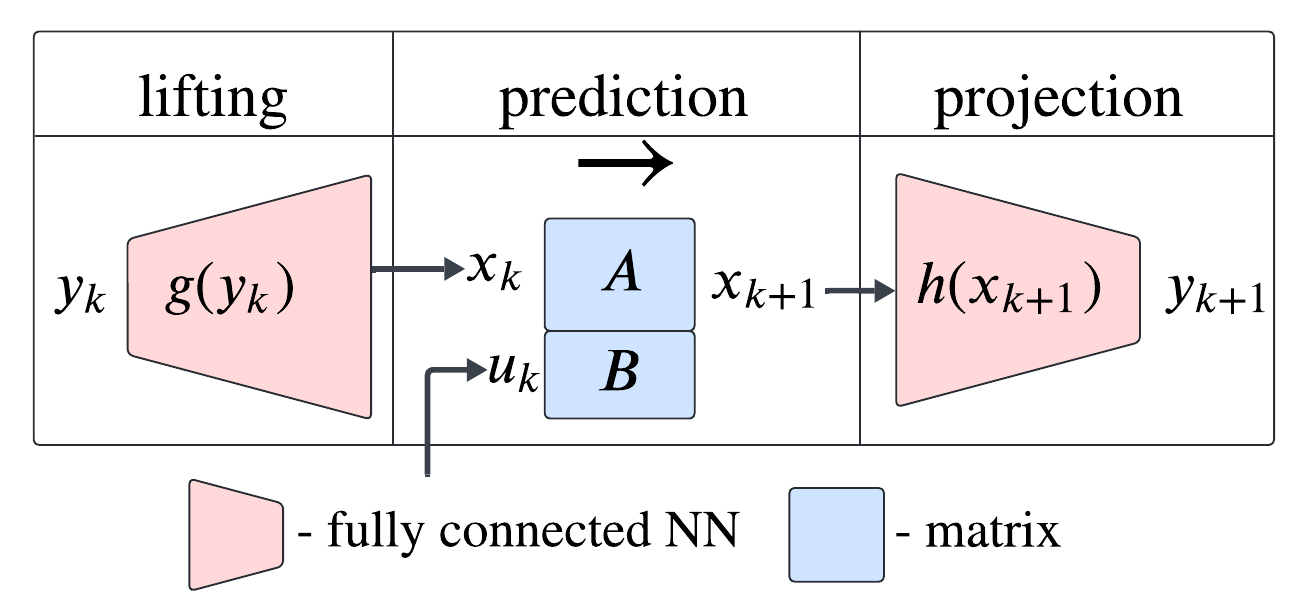}
\caption{Deep Koopman operator with nonlinear projection function.}
\label{fig:deep_koopman_architecture}
\end{figure}

\begin{figure}[h]
\centering
\includegraphics[width=0.8\columnwidth]{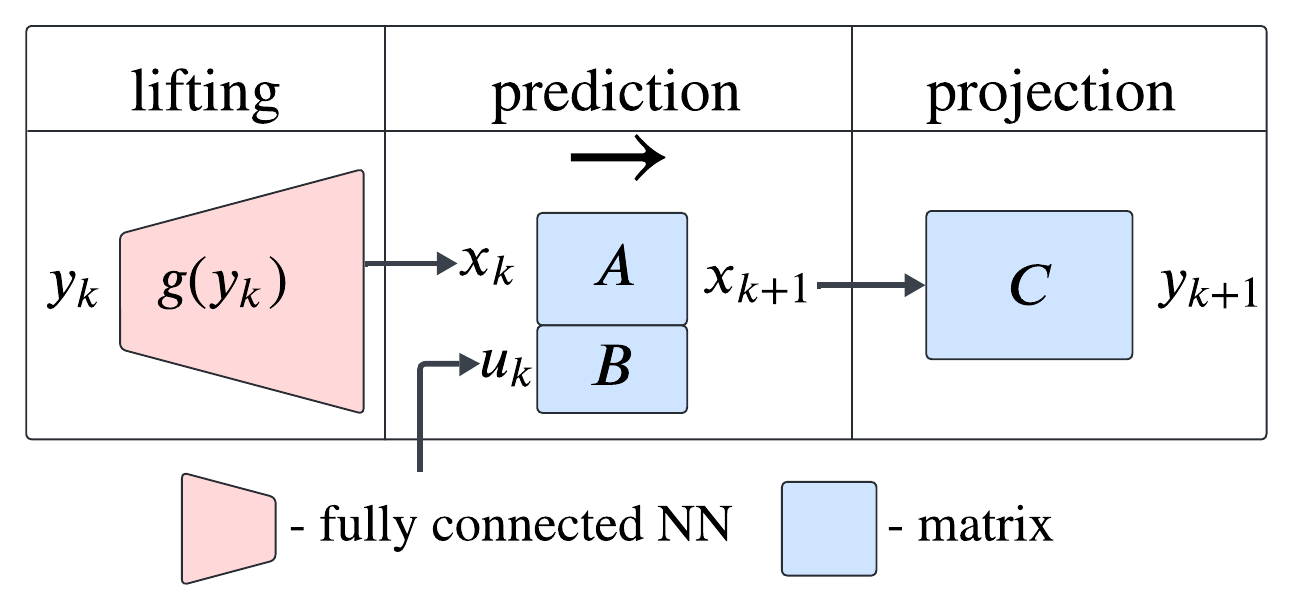}
\caption{Deep Koopman operator with linear projection function.}
\label{fig:deep_koopman_architecture_linear}
\end{figure}

The Koopman operator theory, initially introduced 
by~\cite{koopman1931hamiltonian} and popularized by~\cite{mezic2005spectral}, provides a way to represent nonlinear dynamical systems through linear evolution in a higher-dimensional space. The central idea is to lift the original state space into a lifted space, where the dynamics become linear.
Consider the nonlinear discrete-time system
\begin{equation}
    \label{eq:system}
y_{k+1} = f(y_k, u_k),
\end{equation}
with state $y_k \in \mathbb{R}^{n_{\text{y}}}$, control input $u_k \in \mathbb{R}^{n_{\text{u}}}$, and nonlinear system function $f(\cdot,\cdot)$. The Koopman operator $\mathcal{K}$ acts on lifted functions $g(\cdot): \mathbb{R}^{n_{\text{y}}} \rightarrow \mathbb{R}$ as
$\mathcal{K} g = g \circ f$,
therefore, it evolves linearly in time:
\begin{equation}
\mathcal{K} g(y_k, u_k) = g(f(y_k, u_k)) = g(y_{k+1}).
\end{equation}
The eigenfunctions of $\mathcal{K}$, denoted by $\varphi(y)$, together with corresponding eigenvalues $\lambda$ are the main ingredients in constructing such linear representations, it holds that
$\mathcal{K} \varphi(x) = \lambda \varphi(y),$
which defines invariant directions in the lifted space and allows expressing nonlinear dynamics as linear evolution in terms of these eigenfunctions. However, since $\mathcal{K}$ is infinite-dimensional, obtaining its eigenfunctions analytically is in general infeasible, motivating data-driven approximations.
Since $\mathcal{K}$ is infinite-dimensional, it is approximated using finite-dimensional matrices of a linear-time-invariant system:
\begin{subequations}\label{eq:koopman_predictor}
\begin{align}
x_{k+1} &= A_\mathcal{K} x_k + B_\mathcal{K} u_k,\\
y_k &= C_\mathcal{K} x_k,\\
x_0 &= g(y_0),
\end{align}
\end{subequations}
where $x_k \in \mathbb{R}^{n_x}$ are the lifted states and $A_\mathcal{K} \in \mathbb{R}^{n_x \times n_x}$, $B_\mathcal{K} \in \mathbb{R}^{n_x \times n_u}$, and $C_\mathcal{K} \in \mathbb{R}^{n_y \times n_x}$ are the Koopman matrices.
Data-driven approximations of $\mathcal{K}$ include Dynamic Mode Decomposition (DMD)~\cite{schmid2010dynamic} and its extensions~\cite{williams2015data,proctor2016dynamic,brunton2016discovering}. In this work, we employ a deep learning-based variant known as the Deep Koopman operator
~\cite{lusch2018deep}, which learns both the lifting function and Koopman matrices from data.
This approach allows us to identify the lifting and projection functions without the need to specify them explicitly. The architecture with the included linear dynamics for the control input is shown in Figure~\ref{fig:deep_koopman_architecture}.
However, this approach to modeling cannot be used directly for the linear model predictive control, as the lifting and projection functions are not linear. For the control, the approximation of a projection function with a linear transformation is used, as can be seen in Figure~\ref{fig:deep_koopman_architecture_linear}. The lifting function is used outside the optimization, so it is not necessary to approximate it. Directly using the lifting function can lead to several problems, and a better performance is achieved when using an observer.
\subsection{Economic Model Predictive Control}
Economic Model Predictive Control (EMPC) extends the classical MPC framework by directly optimizing an economic performance index rather than a quadratic tracking objective. The control objective is not merely to follow a reference trajectory, but to operate the system in a way that maximizes efficiency, profitability, or resources. This formulation is particularly suited for energy-intensive systems, where economic considerations dominate tracking problems.

EMPC is defined to predict a discrete-time system in~\eqref{eq:system}
with state $x_k \in \mathbb{R}^{n_{\text{x}}}$, input $u_k \in \mathbb{R}^{n_{\text{u}}}$, within polytopic constraints
\begin{equation}
x_k \in \mathcal{X}, \quad u_k \in \mathcal{U}.
\end{equation}
At each sampling instant $t$, EMPC solves a finite-horizon optimal control problem of the following form with slack variables:
\begin{subequations}\label{eq:empc}
\begin{alignat}{2}
\min_{u_k, \varepsilon_k} \quad &
\sum_{k=0}^{N-1} \big( \ell_{\text{econ}}(x_k, u_k) + c_\varepsilon \varepsilon_k \big)
 \label{eq:empc_cost} \\[1ex]
\text{s.t.} \quad
x_{k+1} &= f(x_k, u_k), \label{eq:empc_dyn} \\[0.5ex]
g(x_k, u_k) &\le \varepsilon_k, \label{eq:empc_constr} \\[0.5ex]
 x_0 &= x(t), \label{eq:empc_init} \\[0.5ex]
 x_k &\in \mathcal{X}, \\
u_k &\in \mathcal{U}, \\
\varepsilon_k &\ge 0,\\
 k &= 0, \dots, N-1. \label{eq:empc_sets}
\end{alignat}
\end{subequations}
Here, $\ell_{\text{econ}}(x_k, u_k)$ denotes the economic stage cost at prediction step $k$, representing cost related system performance such as energy usage, use of material, or operating cost. The function $g(x_k, u_k)$ represents general constraints of the system, which are softened by the slack variables to maintain feasibility. Much like the tracking or regulatory MPC, the stage cost is computed over the prediction horizon $N$. Unlike the quadratic cost typically used in tracking MPC, $\ell(x, u)$ does not have to be positive definite, allowing the controller to explore operating conditions that are economically favorable.
The inclusion of slack variables $\varepsilon_k$ serves to relax hard constraints, ensuring feasibility even under disturbances, model mismatch, or actuator limitations. The term $c_\varepsilon \varepsilon_k$ penalizes constraint violations and ensures satisfaction of physical or safety limits. A sufficiently large weighting factor $c_\varepsilon$ discourages violations, but preserves feasibility in cases where constraints would otherwise cause an infeasible solution. Depending on the application, quadratic or linear penalties on $\varepsilon_k$ can be adopted to shape the relaxation.

\section{Pasteurization Plant}
\label{sec:pasteurization_plant}
\begin{figure}
  \centering
  \includegraphics[width=0.35\textwidth]{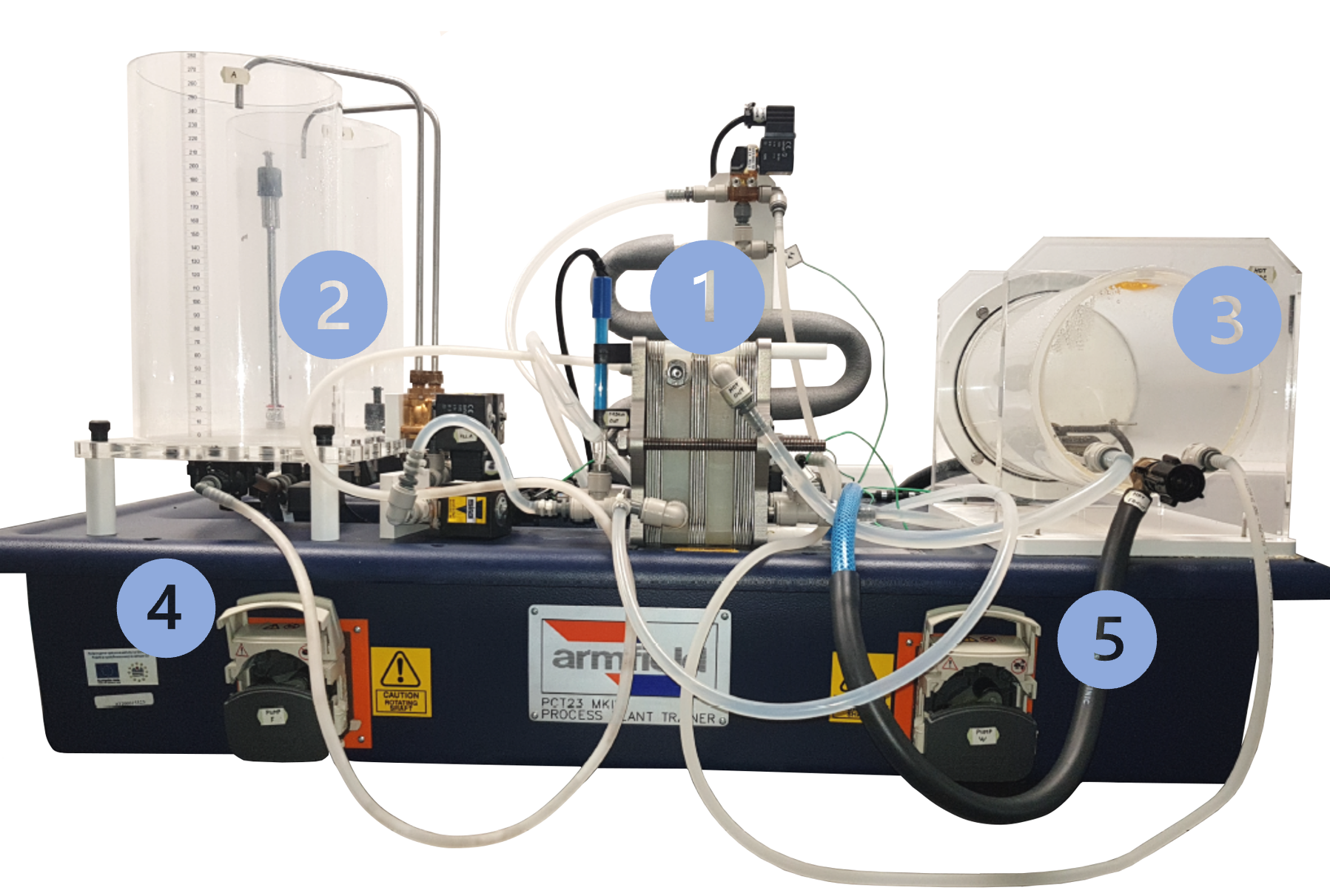}
  \caption{Laboratory pasteurization unit: (1) plate heat exchanger, 
  (2) cold medium tank, (3) hot medium tank with heater, (4) cold feed pump, (5) hot medium pump.}
  \label{fig:pu-foto}
\end{figure}

\begin{figure}
  \centering
  \includegraphics[width=0.45\textwidth]{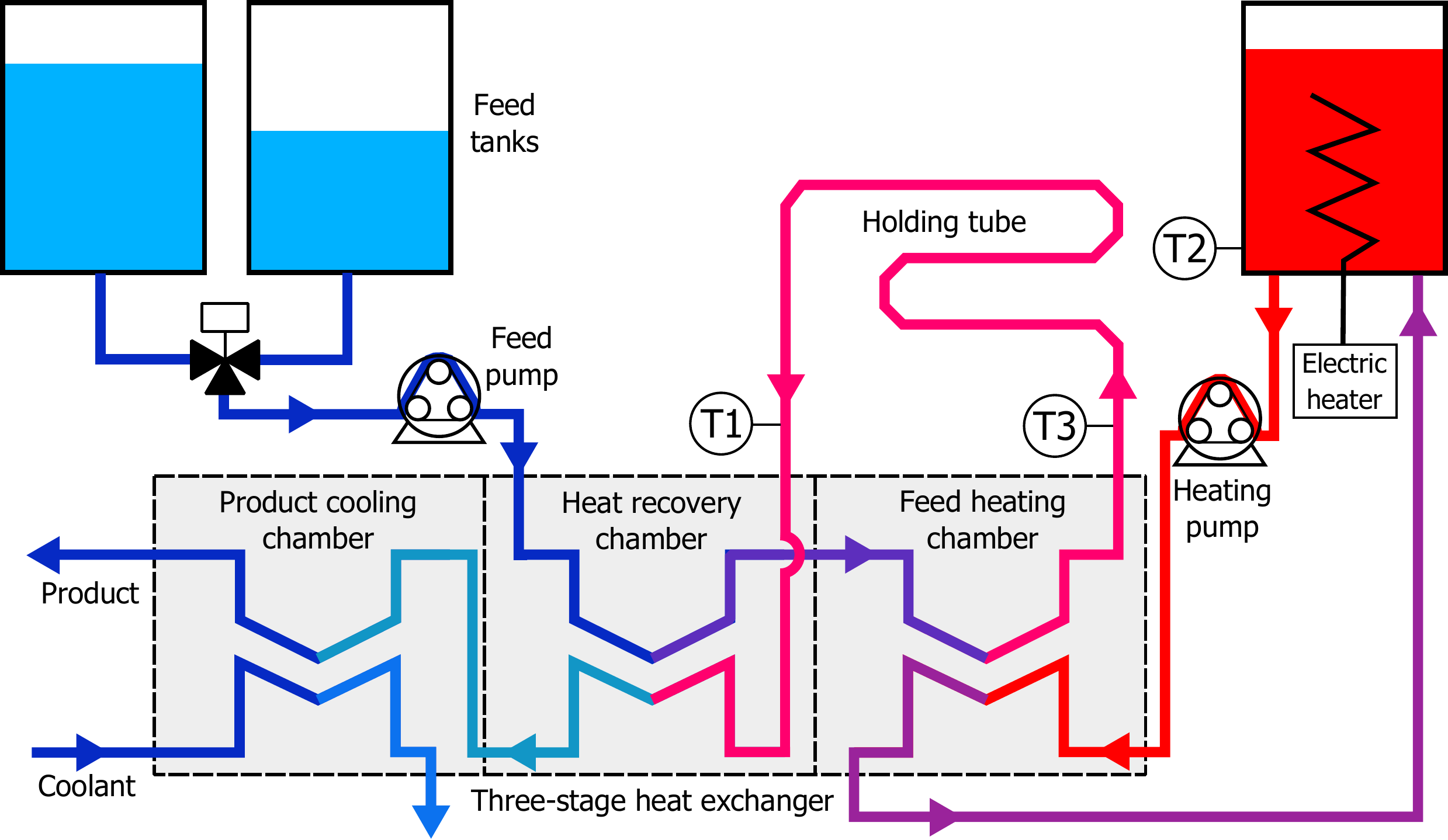}
  \caption{Process scheme of the pasteurization unit. 
  Variables: \(u_1\) – feed pump flow rate, \(u_2\) – heating pump flow rate, \(u_3\) – heater power, 
  \(y_1\) – outlet temperature $T_1$, \(y_2\) – tank temperature $T_2$, \(y_3\) – outlet temperature $T_3$.}
  \label{fig:pu-scheme}
\end{figure}

The experimental setup is a laboratory-scale pasteurization unit (PU) shown in Figure~\ref{fig:pu-foto}. 
This benchmark represents an energy-intensive multidimensional process with three manipulated variables and three controlled outputs, 
serving as a compact version of an industrial pasteurization line. The plant layout is depicted in Figure~\ref{fig:pu-scheme}.

The PU consists of a plate heat exchanger with heat recovery, product heating, and feed cooling zones. 
The cold product from a feed tank is pumped through the heat recovery section, where it absorbs residual energy from already pasteurized liquid. The product then enters the heating chamber, where it is heated by a circulating hot medium. The temperature of the hot water is maintained by the electrical heater with a power input. The manipulated variables are: \(u_1\) flow rate of the cold medium $q_{\text{cold}}$, \(u_2\) hot medium with flow rate $q_{\text{hot}}$,  \(u_3\) electrical heater with power input $P$. 
The corresponding controlled variables are: \(y_1\) the outlet temperature at the holding tube $T_1$, 
\(y_2\) the temperature in the heating tank $T_2$, and \(y_3\) the temperature at the heat exchanger outlet $T_3$. 
The control objective is to maintain the required pasteurization temperature while minimizing energy consumption and product loss.
Before model identification, all input and output signals were standardized using the 
\texttt{StandardScaler} from the \texttt{scikit-learn} library.

\section{Economic Model Predictive Control of Pasteurization Unit}
The Economic Model Predictive Control (EMPC) formulation employed in this work optimizes an economic cost function while ensuring constraint satisfaction over a finite prediction horizon.
 The coefficients in the cost function were derived directly from physical economic considerations and are summarized in Table~\ref{tab:cost_weights}. Here, $Q_u = 4.56 \in I^{3}$, i.e., an identity matrix scaled by the given coefficient $4.56$. All cost values are expressed in euro cents.  Only the first optimal control action $u_0^*$ is applied to the system, and the following optimization is repeated at each sampling instant:

\begin{subequations}
\label{eq:empc_pu}
\begin{align}
\min_{u,\varepsilon_1,\varepsilon_3,\varepsilon_1^{\text{s}},\varepsilon_3^{\text{s}},\varepsilon_1^{\text{u}}} \;
& J(\Delta u, u, \varepsilon_1, \varepsilon_3, \varepsilon_1^{\text{s}}, \varepsilon_3^{\text{s}}, \varepsilon_1^{\text{u}}) \label{eq:empc:obj} \\[1mm]
\text{s.t.} \quad
x_{k+1} &= A x_k + B u_k, \label{eq:empc:state}\\
y_k &= C x_k + d_k, \label{eq:empc:output}\\
\Delta u_k &= u_k - u_{k-1}, \label{eq:empc:deltau}\\
u_{\min,1}-\varepsilon_{k,1}^{\text{u}} &\le u_{k,1} \le u_{\max,1}+\varepsilon_{k,1}^{\text{u}}, \label{eq:empc:u1}\\
u_{\min,j} &\le u_{k,j}\le u_{\max,j},\; j\in\{2,3\}, \label{eq:empc:uj}\\
y_{\min,i}-\varepsilon_{k,i}-\varepsilon_{k,i}^{\text{s}} &\le y_{k,i}\le y_{\max,i},\; i\in\{1,3\}, \label{eq:empc:ysoft}\\
y_{\min,2} &\le y_{k,2}\le y_{\max,2}, \label{eq:empc:yhard}\\
0 &\le \varepsilon_{k,i}\le \varepsilon_{\max,i},\; i\in\{1,3\}, \label{eq:empc:eps1}\\
0 &\le \varepsilon_{k,i}^{\text{s}}\le \varepsilon_{\max,i}^{\text{s}},\; i\in\{1,3\}, \label{eq:empc:eps2}\\
0 &\le \varepsilon_{k,1}^{\text{u}}\le \varepsilon_{\max,1}^{\text{u}}, \label{eq:empc:epsu}\\
x_0 &= \hat{x}(t), \label{eq:empc:initx}\\
d_k &= \hat{d}(t), \label{eq:empc:initd}\\
u_{-1} &= u(t-\Ts), \label{eq:empc:uprev}\\
k &= 0,\ldots,N\!-\!1. \label{eq:empc:k}
\end{align}
\end{subequations}
where $x_k$ denotes the system state, $u_k$ is the control input, $y_k$ is the system output, $d_k$ is the disturbance. 
The matrices $A$, $B$, and $C$ define the system dynamics~\eqref{eq:empc:state} and~\eqref{eq:empc:output}. 
The initial state estimate $\hat{x}(t)$~\eqref{eq:empc:initx}, and disturbance estimation $d=\hat{d}(t)$~\eqref{eq:empc:initd} are provided by a Kalman filter. 
The constraint~\eqref{eq:empc:u1} includes slack variables $\varepsilon_{k,1}^{\text{u}}$ that allow temporary violations of the first input bounds, 
while the input limits for  $u_j,~~ j\in\{2,3\}$ are enforced as hard constraints in~\eqref{eq:empc:uj}. 
Soft output bounds for $y_i,~~ i\in\{1,3\}$ are relaxed using the slack variables $\varepsilon_{k,i}$ and $\varepsilon_{k,i}^{\text{s}}$ in~\eqref{eq:empc:ysoft}, 
whereas the second output $y_{k,2}$ remains hard-constrained according to~\eqref{eq:empc:yhard}. 
The admissible ranges of all slack variables are restricted by~\eqref{eq:empc:eps1}--\eqref{eq:empc:epsu}.

\subsection{Economic Cost function of PU}
The economic objective function $J$ is defined as:
\begin{equation}
\begin{aligned}
J = \sum_{k=0}^{N-1} \Big( & c_{\text{energy}} u_{k,3} + c_{\text{material}} (\varepsilon_{k,1} + \varepsilon_{k,3})
 + c_{\text{input}} \varepsilon_{k,1}^{\text{u}}+ \\&+ c_{\text{soft}} (\varepsilon_{k,1}^{\text{s}} + \varepsilon_{k,3}^{\text{s}}) 
 + \Delta u_k^\top Q_u \Delta u_k \Big),
\end{aligned}
\end{equation}
where the cost function terms are designed to balance economic performance and operational constraints:
\begin{itemize}
    \item Energy cost penalized by $c_{\text{energy}}$, the energy cost coefficient, representing the economic cost of control effort. This constant is set as the cost of electrical energy for the operation of the heating spiral \(u_3\), determined by the electricity tariff applicable to the PU.
    \item Material loss penalized by $c_{\text{material}}$, the penalty for material losses due to output constraint violations. The product is below the safe temperature of pausterization and has to be discarded. The penalty is applied for both outputs (temperatures) \(y_1\) and \(y_3\) because in either case the product is not pasteurized and has to be discarded. This constant is computed as the price of the product we lost, several times higher than the energy cost. This penalty should be computed as a mixed integer problem, but to keep the problem convex, it is approximated by continuous slack variables.
    \item Soft input slack penalized by $c_{\text{input}}$, the penalty for input constraint violations. This penalty is for the desired range of the flow rate \(u_1\) to be maintained, but not limiting the full capacity of the pump if needed. The penalty should be higher than the energy cost to ensure that in the steady-state, the input is at the desired range.
    \item Soft output slacks penalized by the $c_{\text{soft}}$, low penalty for outputs nearing the constraints of the pasteurization temperature. This penalty is applied for both outputs \(y_1\) and \(y_3\). Thanks to this penalization, the controller avoids nearing the output constraints. 
    \item Input movement penalized by $Q_{\text{u}}$, the penalty matrix for input rate of change $\Delta u_k = u_k - u_{k-1}$, ensuring smooth control actions, avoiding bang-bang control and actuator wear. 
\end{itemize}

\begin{table}
\centering
\caption{EMPC objective function weights and coefficients. All scalar cost coefficients are expressed in euro cents.}
\label{tab:cost_weights}
\begin{tabular}{lll}
\toprule
Coefficient & Value & Ratio to $c_{\text{energy}}$ \\
\midrule
$c_{\text{energy}}$ & $4.56 \cdot 10^{-2}$ & $1$ \\
$c_{\text{material}}$ & $4.78 $ & $105$ \\
$c_{\text{input}}$ & $4.56 \cdot 10^{-1}$ & $10$ \\
$c_{\text{soft}}$ & $5.92 \cdot 10^{-2}$ & $1.3$ \\
$Q_{\text{u}}$ & $4.56 \in I^{3}$ & $100$ \\

\bottomrule
\end{tabular}
\end{table}
\subsection{Constraints of PU}
The control problem is subject to both input and output constraints. Input constraints represent physical limitations of the actuators (flow rates and heater power), while output constraints ensure that the pasteurization temperatures remain within safe operating limits. The specific limit values are presented in Table~\ref{tab:constraints}.

\begin{table}
\centering
\caption{Input and Output Constraints (Descaled Physical Units)}
\label{tab:constraints}
\begin{tabular}{lccl}
\toprule
Inputs & $u_{\text{min}}$ & $u_{\text{max}}$ & Unit \\
\midrule 
$u_1$ (soft) & 5.8 & 7.4 & \si{\cubic\centi\meter\per\second} \\
$u_1$ (hard) & 1.3 & 11.9 & \si{\cubic\centi\meter\per\second} \\
$u_2$ & 0.8 & 11.5 & \si{\cubic\centi\meter\per\second} \\
$u_3$ & 0.25 & 1 & \si{\kilo\watt} \\
\midrule
Outputs & $y_{\text{min}}$ & $y_{\text{max}}$ & Unit \\
\midrule  
$y_1$  & 72.5 & 100 & \si{\celsius} \\
$y_2$  & 0 & 100 & \si{\celsius} \\
$y_3$  & 73.0 & 100 & \si{\celsius} \\
\midrule
Slacks & $\varepsilon_{\text{min}}$ & $\varepsilon_{\text{max}}$ & Unit \\
\midrule
$\varepsilon_1$, $\varepsilon_3$ & 0 & $\infty$ & \si{\celsius} \\[1mm]
$\varepsilon_1^{\text{s}}$, $\varepsilon_3^{\text{s}}$ & 0 & 0.5 & \si{\celsius} \\[1mm]
$\varepsilon_1^{\text{u}}$ & 0 & 4.5 & \si{\cubic\centi\meter\per\second} \\
\bottomrule
\end{tabular}
\end{table}

To guide the feed flow rate \(u_1\) into the desired range, but not limiting the full capacity of the actuator pump, we used a slack variable \(\varepsilon^{\text{u}}\) to allow temporary violations of the input constraints. Other input slack variables were not used as they are not needed.

To handle constraint violations during disturbances and fault scenarios, two types of output slack variables were introduced: (i) highly penalized output slack $\varepsilon_{k,1}$ and $\varepsilon_{k,3}$, that permits violations of output lower constraints with high penalty (the product is not pasteurized and has to be discarded).  (ii) softly penalized output slack $\varepsilon^{\text{s}}_{k,1}$ and $\varepsilon^{\text{s}}_{k,3}$ set to penalize the pasteurization lower bound temperatures. Output slack variables $\varepsilon_{k,1}$ and $\varepsilon^{\text{s}}_{k,1}$ penalize output \(y_1\), while variables $\varepsilon_{k,3}$ and $\varepsilon^{\text{s}}_{k,3}$ penalize output \(y_3\). The specific limit values of all slacks $\varepsilon_{\text{min}}$ and $\varepsilon_{\text{max}}$ are presented in Table~\ref{tab:constraints}.

\section{Identification Setup}
The three models were identified in this work. The Deep Koopman model with a nonlinear projection function is used as a simulation model, as it provides the highest accuracy. However, it cannot be used directly for the linear EMPC, since the projection function is nonlinear. Therefore, a different Deep Koopman model with a linear projection function is trained for the MPC prediction to meet controllability and observability requirements. Both models are implemented in the \textsc{Neuromancer} library~\cite{Neuromancer2023} in Python. N4SID model is used as a reference model, as it represents a widely used model in the literature and is a good baseline for the comparison. 

All models were identified from the same experimental data and evaluated on the same test dataset, which was collected from the real plant described in Section~\ref{sec:pasteurization_plant}. Both datasets were created during separate experiments, however, different step changes of the control were applied to the system to generate the data. Both datasets contain about \(30 \si{\minute}\) of data collected.

\begin{figure}
\centering
\includegraphics[width=0.9\columnwidth]{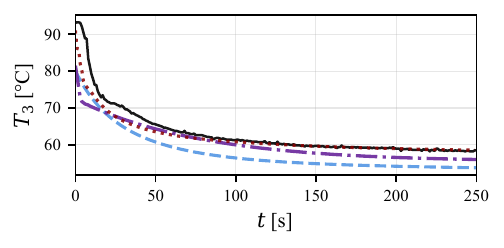}
\caption{Open-loop identification validation: measured data (black) vs. N4SID (\textcolor{n4sidblue}{blue} dashed), Deep Koopman linear (\textcolor{koopmanpurple}{purple} dash-dot), and Deep Koopman nonlinear (\textcolor{koopmanred}{dark red} dotted) for output \(T_3\).}
\label{fig:identification_comparison}
\end{figure}

The comparison of the identification performance of all three models is shown in Table~\ref{tab:model_comparison} and Figure~\ref{fig:identification_comparison}. As we can see, the Deep Koopman model with a nonlinear projection function achieves the lowest mean absolute error across all outputs, with an improvement of $93\%$ compared to the N4SID baseline. The Deep Koopman model with a linear projection function still achieves a lower mean absolute error than the N4SID baseline, with an improvement of $45\%$. These models were identified for the time step \(\Ts = 1\,\si{\second}\) and then rescaled to the time step \(\Ts = 10\,\si{\second}\) for the MPC simulation.

\begin{table}
    \centering
    \caption{Identification performance comparison: Mean Absolute Error (MAE) across all outputs on test dataset. All values in \si{\celsius}.}
    \label{tab:model_comparison}
    \begin{tabular}{lcc}
    \toprule
    Model & MAE & Improvement \\
    \midrule
    N4SID Baseline & 0.763 & --- \\
    Deep Koopman (nonlinear \(h(x)\)) & 0.055 & $93\%$ \\
    Deep Koopman (linear $Cx$) & 0.421 & $45\%$ \\
    \bottomrule
    \end{tabular}
    \end{table}

\subsection{Deep Koopman nonlinear control model}
The Deep Koopman model lifts the system to a \(30\)-dimensional latent space, chosen arbitrarily as \(10 \cdot n_y\). The encoder $g(y_k)$ is a 3-layer MLP with hidden layer dimensions
\(
\begin{bmatrix}
60 & 120 & 180
\end{bmatrix}
\)
and ReLU activations, mapping the \(3\)-dimensional output to the \(30\)-dimensional lifted state. The Koopman operator $A \in \mathbb{R}^{30 \times 30}$ and input matrix $B \in \mathbb{R}^{30 \times 3}$ govern the linear dynamics in the lifted space. The decoder is a nonlinear 3-layer MLP with hidden layer dimensions
\(
\begin{bmatrix}
180 & 120 & 60
\end{bmatrix}
\)
and ELU activations, mapping back to the \(3\)-dimensional output space.
\subsection{Deep Koopman linear prediction model}
The Deep Koopman model with linear projection function employs a \(26\)-dimensional lifted space. This number is reduced from \(30\) to \(26\) to meet the controllability and observability requirements through iterative methods. The encoder $g(y_k)$ shares the same architecture as the Deep Koopman model and uses hidden layer dimensions
\(
\begin{bmatrix}
60 & 120 & 180
\end{bmatrix}
\)
with ReLU activations. The input encoder is a linear layer mapping the control input to the \(26\)-dimensional space. The key distinction is that the output projection $C \in \mathbb{R}^{3 \times 26}$ is a linear map, enabling exact state-space representation with matrices
\(
A \in \mathbb{R}^{26 \times 26}, \quad B \in \mathbb{R}^{26 \times 3}, \quad C \in \mathbb{R}^{3 \times 26}.
\)
This linearity in $C$ facilitates direct application of linear control techniques.
\subsection{N4SID baseline model }
The N4SID baseline model was identified using MATLAB's System Identification Toolbox (version 2024b) with automatic parameter tuning. The optimal model structure was determined by evaluating mean squared error on the test dataset, yielding a 3-dimensional state-space representation with matrices $A \in \mathbb{R}^{3 \times 3}$, $B \in \mathbb{R}^{3 \times 3}$, and $C \in \mathbb{R}^{3 \times 3}$. This model serves as a benchmark representing conventional subspace identification methods~\cite{van2012subspace} widely employed in industrial practice.

\section{Control Setup}
\label{sec:control_setup}

This section describes the complete control setup used for both the N4SID and Koopman-based EMPC implementations. All simulations were conducted under identical conditions described in this section to ensure fair comparison.
%\subsection{Simulation parameters}
The closed-loop simulations were performed over a total time span of \(1440\) steps, corresponding to \(4\) hours of operation with a sampling time \(\Ts = 10\,\si{\second}\). The MPC optimization horizon was set to \(N = 60\) steps, providing a 10-minute prediction window, which is enough to capture the system dynamics. Both controllers used GUROBI~\cite{gurobi} as the quadratic programming solver.
%\subsection{State estimation}
A Kalman filter~\cite{kalman1960new} was used for state estimation and disturbance observation. The state vector \(\hat{x}_k\) was augmented with output disturbances \(d_k \in \mathbb{R}^{n_{\text{y}}}\) to mitigate the plant model mismatch and improve the control performance. The Kalman filter parameter \(\ui{Q}{kf} \in \mathbb{R}^{(n_{\text{x}}+n_{\text{y}})\times (n_{\text{x}}+n_{\text{y}})}\) is an identity matrix scaled by \(0.1\) and \(\ui{R}{kf} \in \mathbb{R}^{n_{\text{y}}\times n_{\text{y}}}\) is an identity matrix scaled by \(0.5\) were tuned to minimize the estimation error and disturbance observation error. The initial error covariance \(\ui{P}{0,kf} \in \mathbb{R}^{(n_{\text{x}}+n_{\text{y}})\times (n_{\text{x}}+n_{\text{y}})}\) is an identity matrix.
The matrices are identical for both models with corresponding dimensions of the state vector.
%\subsection{External disturbance}
To evaluate controller robustness, two disturbance scenarios were simulated. The first disturbance occurs at \(t = 0\): here, the system is initialized with output temperatures $T_1$ and $T_3$ set below their minimum constraints, simulating the introduction of a new batch of cold pasturized liqiud. The previous input, \(u(t-\Ts)\), was set to the mean value of the control inputs. The second disturbance involves a pump failure in \(t = <720, 780>\), where the feed pump ($u_1$) experiences a complete control signal loss for 60 time steps (10 minutes), forcing it to remain at its minimum constraint value in a fail-to-close scenario. During the pump failure period, the input slack penalty for $u_1$ was excluded from the cost function to avoid penalizing the controller for pump failures outside of its control.

\section{Results}

\begin{figure}
    \centering
    \includegraphics[width=0.9\columnwidth]{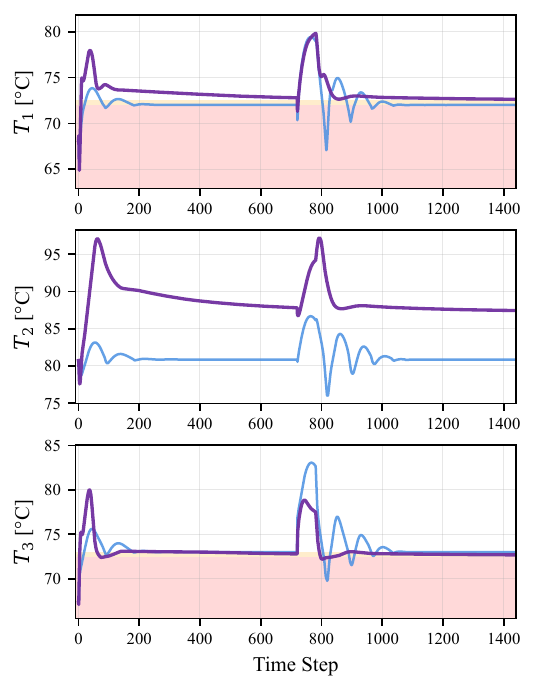}
    \caption{Output trajectories: Koopman-based EMPC (\textcolor{koopmanpurple}{purple}) with linear projection vs. N4SID EMPC (\textcolor{n4sidblue}{blue}). \textcolor{constraintred}{Red} zone indicates inadequate pasteurization temperature; \textcolor{softorange}{orange} zone shows soft constraint region.}
    \label{fig:outputs}
    \end{figure}

\begin{figure}
    \centering
    \includegraphics[width=0.9\columnwidth]{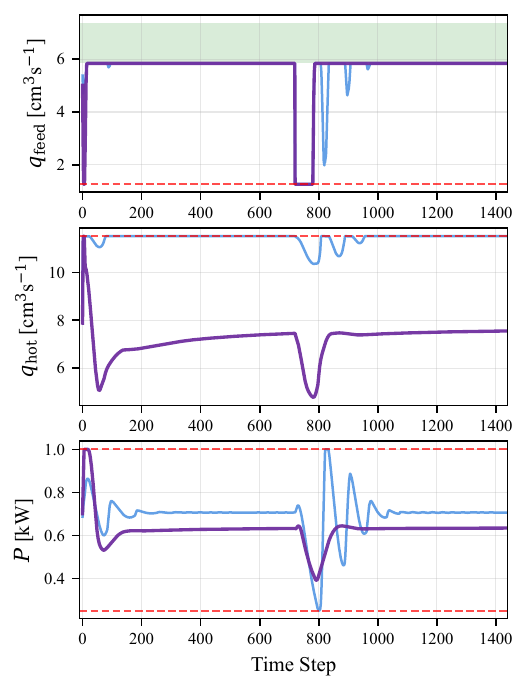}
    \caption{Input trajectories: Koopman-based EMPC (\textcolor{koopmanpurple}{purple}) with linear projection vs. N4SID EMPC (\textcolor{n4sidblue}{blue}). \textcolor{constraintred}{Red} dashed lines show hard constraints; \textcolor{desiredgreen}{green} area shows desired range for \(\ui{q}{feed}\).}
    \label{fig:inputs}
    \end{figure}

In this section, we present and analyze the simulation results comparing the closed-loop performance of the EMPC controller using the N4SID and the deep Koopman model. The simulations are presented in Figure~\ref{fig:outputs}, Figure~\ref{fig:inputs}. Both controllers were subjected to identical operating conditions, disturbances and control setup described in~\ref{sec:control_setup}.
The performance evaluation is based on the economic objective function defined in \ref{eq:empc:obj}. Table~\ref{tab:cost_comparison} summarizes the closed-loop cost analysis for both considered models. The improvement factor is computed as the ratio of the N4SID cost to the Koopman cost. %The individual components of the cost function are also visualized in Figure~\ref{fig:cost}. 
\begin{table}[t]
\centering
\caption{Closed-loop cost analysis of N4SID and Koopman-based EMPC. }
\label{tab:cost_comparison}
\begin{tabular}{lccc}
\hline
Cost Component & N4SID & Koopman & Improvement Factor \\
\hline
Energy Cost (\(u_3\)) & 4.564 & 4.106 & $1.11$ \\
Material Loss (\(T_1\)) & 1.850 & 0.516 & $3.58$ \\
Material Loss (\(T_3\)) & 0.890 & 0.283 & $3.15$ \\
Soft Constraint (\(u_1\)) & 0.298 & 0.151 & $1.97$ \\
Soft Constraint (\(T_1\)) & 0.094 & 0.001 & $94.00$ \\
Soft Constraint (\(T_3\)) & 0.006 & 0.035 & $0.17$ \\
Input Movement (\(\Delta u\)) & 0.504 & 0.451 & $1.12$ \\
\hline
\textbf{Total Cost} & \textbf{8.207} & \textbf{5.542} & \textbf{$1.48$} \\
\hline
\end{tabular}
\end{table}
    
The results demonstrate that the deep Koopman-based EMPC achieves significantly superior economic performance compared to the N4SID approach. The total closed-loop cost is reduced with an improvement factor of $1.48$, primarily driven by substantial reductions in material losses (improvement factors of $3.58$ for $T_1$ and $3.15$ for $T_3$). These improvements directly translate to reduced product waste in the pasteurization process, as fewer products fail to meet the required temperature specifications. The energy cost of operation is reduced with an improvement factor of $1.11$, indicating that the Koopman-based controller is able to operate the system more efficiently. The input slack constraint for $u_1$ is improved by a factor of $1.97$, and the input movement cost is also reduced with an improvement factor of $1.12$. The only exception is the soft constraint violation for output \(y_3\), which is increased (improvement factor of $0.17$). These soft constraints have an insignificant impact on the total cost and are used only to penalize the close approach to the actual pasteurization lower bound temperatures, but are included for completeness.

Moreover, in the steady-state, the Koopman-based EMPC does consume about \(10.2\%\) less energy than the N4SID EMPC, as it operates in a different operating point than the N4SID EMPC. This is expected as the Koopman EMPC is able to capture the system dynamics better than the N4SID EMPC.
The Koopman-based controller is able to reach and stay above the constraint temperatures faster than the N4SID EMPC.

Overall, the Koopman-based EMPC outperforms the N4SID EMPC across all major economic criteria by $32\%$, demonstrating the practical advantages of the Koopman operator framework for EMPC in energy-intensive processes.

% \begin{figure}[h]
% \centering
% \includegraphics[width=\columnwidth]{figures/cost_comparison.pdf}
% \caption{Comparison of cost components for N4SID and Koopman-based MPC. The Koopman approach achieves significantly lower total cost.}
% \label{fig:cost}
% \end{figure}

\section{CONCLUSIONS}
This work demonstrated that integrating deep Koopman models within Economic MPC improves control performance for the energy-intensive process of pasteurization when compared to the subspace identification method. 
By directly minimizing operational costs, such as energy consumption, material losses, and actuator wear, the deep Koopman-based controller achieved substantial economic improvements over conventional subspace identification methods with an average improvement factor \(1.48\). The framework proved robust under disturbances, including pump failure and cold batch introduction, validating its practical applicability. The methodology requires only input-output data and standard quadratic programming solvers, making it deployable in industrial settings. These improvements on the pasteurization benchmark indicate promising applications for other energy-intensive processes where economic operation is critical. Future work will focus on the experimental implementation of the deep Koopman EMPC on the pasterization unit, including hardware-in-the-loop validation.

\bibliographystyle{IEEEtran}
\bibliography{bibfile}

@article{MAYNE2000789,
title = {Constrained model predictive control: Stability and optimality},
journal = {Automatica},
volume = {36},
number = {6},
pages = {789-814},
year = {2000},
issn = {0005-1098},
doi = {https://doi.org/10.1016/S0005-1098(99)00214-9},
author = {D.Q. Mayne and J.B. Rawlings and C.V. Rao and P.O.M. Scokaert}
}

@INPROCEEDINGS{rawlings-empc,
  author={Rawlings, James B. and Angeli, David and Bates, Cuyler N.},
  booktitle={2012 IEEE 51st IEEE Conference on Decision and Control (CDC)}, 
  title={Fundamentals of economic model predictive control}, 
  year={2012},
  volume={},
  number={},
  pages={3851-3861},
  doi={10.1109/CDC.2012.6425822}}

@article{ellis2017economic,
  title={Economic model predictive control},
  author={Ellis, Matthew and Liu, Jinfeng and Christofides, Panagiotis D},
  journal={Springer},
  volume={5},
  number={7},
  pages={65},
  year={2017},
  publisher={Springer}
}

@article{empc-cstr,
title = {End-to-end reinforcement learning of Koopman models for economic nonlinear model predictive control},
journal = {Computers \& Chemical Engineering},
volume = {190},
pages = {108824},
year = {2024},
issn = {0098-1354},
doi = {https://doi.org/10.1016/j.compchemeng.2024.108824},
author = {Daniel Mayfrank and Alexander Mitsos and Manuel Dahmen}
}

@ARTICLE{empc-wave,
  author={Jia, Yubin and Sun, Jia and Xu, Zhao and Sun, Changyin and Meng, Ke and Dong, Zhao Yang},
  journal={IEEE Journal of Emerging and Selected Topics in Industrial Electronics}, 
  title={Data-Driven Economic MPC of a Point Absorber Wave Energy Converter}, 
  year={2024},
  volume={5},
  number={2},
  pages={670-679},
  doi={10.1109/JESTIE.2024.3363668}}

@INPROCEEDINGS{empc-ded,
  author={King, Ethan and Drgo\v{n}a, Ján and Tuor, Aaron and Abhyankar, Shrirang and Bakker, Craig and Bhattacharya, Arnab and Vrabie, Draguna},
  booktitle={2022 American Control Conference (ACC)}, 
  title={Koopman-based Differentiable Predictive Control for the Dynamics-Aware Economic Dispatch Problem}, 
  year={2022},
  volume={},
  number={},
  pages={2194-2201},
  doi={10.23919/ACC53348.2022.9867379}}

@article{empc-wastewater,
title = {Efficient economic model predictive control of water treatment process with learning-based Koopman operator},
journal = {Control Engineering Practice},
volume = {149},
pages = {105975},
year = {2024},
issn = {0967-0661},
doi = {https://doi.org/10.1016/j.conengprac.2024.105975},
author = {Minghao Han and Jingshi Yao and Adrian Wing-Keung Law and Xunyuan Yin}
}

@article{brunton2016discovering,
  title     = {Discovering governing equations from data by sparse identification of nonlinear dynamical systems},
  author    = {Brunton, Steven L and Proctor, Joshua L and Kutz, J Nathan},
  journal   = {Proceedings of the national academy of sciences},
  volume    = {113},
  number    = {15},
  pages     = {3932--3937},
  year      = {2016},
  publisher = {National Academy of Sciences}
}

@inproceedings{valabek2025deep,
  title        = {Deep Dictionary-Free Method for Identifying Linear Model of Nonlinear System with Input Delay},
  author       = {Val{\'a}bek, Patrik and Wadinger, Marek and Kvasnica, Michal and Klau\v{c}o, Martin},
  booktitle    = {2025 25th International Conference on Process Control (PC)},
  pages        = {1--6},
  year         = {2025},
  organization = {IEEE}
}

@book{van2012subspace,
  title     = {Subspace identification for linear systems: Theory—Implementation—Applications},
  author    = {Van Overschee, Peter and De Moor, BL0888},
  year      = {2012},
  publisher = {Springer Science \& Business Media}
}

@article{kalman1960new,
  author  = {Kalman, Rudolph Emil},
  title   = {A New Approach to Linear Filtering and Prediction Problems},
  journal = {Transactions of the ASME--Journal of Basic Engineering},
  volume  = {82},
  number  = {Series D},
  pages   = {35--45},
  year    = {1960}
}

@misc{gurobi,
  author = {{Gurobi Optimization, LLC}},
  title  = {{Gurobi Optimizer Reference Manual}},
  year   = 2024,
  url    = {https://www.gurobi.com}
}

@article{korda2020optimal,
  title     = {Optimal construction of Koopman eigenfunctions for prediction and control},
  author    = {Korda, Milan and Mezi{\'c}, Igor},
  journal   = {IEEE Transactions on Automatic Control},
  volume    = {65},
  number    = {12},
  pages     = {5114--5129},
  year      = {2020},
  publisher = {IEEE}
}

@article{Neuromancer2023,
  title  = {{NeuroMANCER: Neural Modules with Adaptive Nonlinear Constraints and Efficient Regularizations}},
  author = {Drgo{\v{n}}a, Jan and Tuor, Aaron and Koch, James and Shapiro, Madelyn and Vrabie, Draguna},
  url    = {https://github.com/pnnl/neuromancer},
  year   = {2023}
}

@article{schmid2010dynamic,
  title     = {Dynamic mode decomposition of numerical and experimental data},
  author    = {Schmid, Peter J},
  journal   = {Journal of fluid mechanics},
  volume    = {656},
  pages     = {5--28},
  year      = {2010},
  publisher = {Cambridge University Press}
}

@article{korda2018linear,
  title     = {Linear predictors for nonlinear dynamical systems: Koopman operator meets model predictive control},
  author    = {Korda, Milan and Mezi{\'c}, Igor},
  journal   = {Automatica},
  volume    = {93},
  pages     = {149--160},
  year      = {2018},
  publisher = {Elsevier}
}

@article{proctor2016dynamic,
  title     = {Dynamic mode decomposition with control},
  author    = {Proctor, Joshua L and Brunton, Steven L and Kutz, J Nathan},
  journal   = {SIAM Journal on Applied Dynamical Systems},
  volume    = {15},
  number    = {1},
  pages     = {142--161},
  year      = {2016},
  publisher = {SIAM}
}

@inproceedings{yeung2019learning,
  title        = {Learning deep neural network representations for Koopman operators of nonlinear dynamical systems},
  author       = {Yeung, Enoch and Kundu, Soumya and Hodas, Nathan},
  booktitle    = {2019 American Control Conference (ACC)},
  pages        = {4832--4839},
  year         = {2019},
  organization = {IEEE}
}

@article{lusch2018deep,
  title     = {Deep learning for universal linear embeddings of nonlinear dynamics},
  author    = {Lusch, Bethany and Kutz, J Nathan and Brunton, Steven L},
  journal   = {Nature communications},
  volume    = {9},
  number    = {1},
  pages     = {4950},
  year      = {2018},
  publisher = {Nature Publishing Group UK London}
}

@article{williams2015data,
  title     = {A data--driven approximation of the koopman operator: Extending dynamic mode decomposition},
  author    = {Williams, Matthew O and Kevrekidis, Ioannis G and Rowley, Clarence W},
  journal   = {Journal of Nonlinear Science},
  volume    = {25},
  pages     = {1307--1346},
  year      = {2015},
  publisher = {Springer}
}

@article{mezic2005spectral,
  title     = {Spectral properties of dynamical systems, model reduction and decompositions},
  author    = {Mezi{\'c}, Igor},
  journal   = {Nonlinear Dynamics},
  volume    = {41},
  pages     = {309--325},
  year      = {2005},
  publisher = {Springer}
}

@article{koopman1931hamiltonian,
  title     = {Hamiltonian systems and transformation in Hilbert space},
  author    = {Koopman, Bernard O},
  journal   = {Proceedings of the National Academy of Sciences},
  volume    = {17},
  number    = {5},
  pages     = {315--318},
  year      = {1931},
  publisher = {National Acad Sciences}
}

@article{MARTIN2018,
  title   = {Symposium review: Effect of post-pasteurization contamination on fluid milk quality},
  journal = {Journal of Dairy Science},
  volume  = {101},
  number  = {1},
  pages   = {861-870},
  year    = {2018},
  issn    = {0022-0302},
  doi     = {https://doi.org/10.3168/jds.2017-13339},
  author  = {Nicole H. Martin and Kathryn J. Boor and Martin Wiedmann}
}

\end{document}